\documentclass[preprint,showpacs,amsmath,amssymb,nofootinbib]{revtex4}
\usepackage{changes}
\usepackage{amsfonts}
\usepackage{amsmath,graphicx,color,epsfig}

\newcommand{\be}{\begin{equation}}
\newcommand{\ee}{\end{equation}}
\newcommand{\bea}{\begin{eqnarray}}
\newcommand{\eea}{\end{eqnarray}}
\newcommand{\ba}{\begin{array}}
\newcommand{\ea}{\end{array}}
\newcommand{\bi}{\begin{itemize}}
\newcommand{\ei}{\end{itemize}}

\newcommand{\nslash}{\kern 0.2 em n\kern -0.50em /}
\newcommand{\kslash}{\kern 0.2 em k\kern -0.45em /}
\newcommand{\qslash}{\kern 0.2 em q\kern -0.45em /}
\newcommand{\pslash}{\kern 0.2 em p\kern -0.50em /}
\newcommand{\rslash}{\kern 0.2 em r\kern -0.50em /}
\newcommand{\sslash}{\kern 0.2 em s\kern -0.50em /}
\newcommand{\Sslash}{\kern 0.2 em S\kern -0.50em /}
\newcommand{\Pslash}{\kern 0.2 em P\kern -0.50em /}
\newcommand{\Dslash}{\kern 0.2 em D\kern -0.65em /\kern 0.15em}
\newcommand{\lf}{\left}
\newcommand{\rg}{\right}

\begin{document}
\title{Vector mesons and electromagnetic form factor of the $\Lambda$ hyperon}

\author{Xu Cao}
\affiliation{CAS Key Laboratory of High Precision Nuclear Spectroscopy and Center for Nuclear Matter Science, Institute
of Modern Physics, Chinese Academy of Sciences, Lanzhou 730000, China}
\affiliation{
State Key Laboratory of Theoretical Physics, Institute of Theoretical Physics, Chinese Academy of Sciences, Beijing 100190,
China}

\author{Jian-Ping Dai{\footnote{Corresponding author: daijianping@ihep.ac.cn}}}
\affiliation{INPAC, Shanghai Key Laboratory for Particle Physics and Cosmology, MOE Key Laboratory for Particle Physics,
Astrophysics and Cosmology, Shanghai Jiao-Tong University, Shanghai, 200240, China}
\affiliation{
State Key Laboratory of Theoretical Physics, Institute of Theoretical Physics, Chinese Academy of Sciences, Beijing 100190,
China}

\author{Ya-Ping Xie}
\affiliation{Institute of Modern Physics, Chinese Academy of Sciences, Lanzhou 730000, China}

\begin{abstract}
The measured electromagnetic form factors of $\Lambda$ hyperon in the time-like region are significantly deviated from pQCD
prediction. We attribute the non-vanishing cross section near threshold to be the contribution of below-threshold $\phi$(2170)
state, supporting its exotic structure. Above the threshold, we find significant role of a wide vector meson with the mass
of around 2.34 GeV, which would be the same state present in $p\bar{p}$ annihilation reactions. As a result, we give a
satisfactory description of the behavior of existing data without modifying pQCD expectation.
\end{abstract}
\maketitle

\section{Introduction} \label{sec:intro}

The electromagnetic form factors (EMFFs) are essential probe of electromagnetic structure of bound states and can deepen
our understanding of perturbative and non-perturbative quantum chromodynamics (QCD) effects encoded in hadrons. The nucleon
EMFFs have been extensively explored from both experimental and theoretical sides for more than sixty years (see reviews~\cite{Denig:2012by,Pacetti:2015iqa}
and references therein). The general behavior of available data~\cite{Lees:2013ebn,Lees:2013uta,Ablikim:2014jrz,Ablikim:2015vga}
in the time-like region tends to be consistent with naive quark counting rules and the pQCD prediction at large-$q^2$~\cite{Lepage:1979za,Belitsky:2002kj}.
It also agrees with the simple extension of the space-like dipole model to the time-like region and most of the nucleon
models, e.g. the constituent quark model~\cite{Bianconi:2015owa,Bianconi:2015vva}. The pQCD predicted that EFFs of other baryons
should follow the same energy dependence. It was only recently that the EMFFs of the $\Lambda$ baryon was measured with good precision~\cite{Bisello:1990rf,Aubert:2007uf,Ablikim:2017pyl,Cui:2018}, so this conclusion could be tested for hyperon for the first time.
However, it seems that the data deviate significantly from the parametrization motivated by pQCD analysis~\cite{Yang:2017hao}.
Besides, a very close-to-threshold enhancement is observed by BESIII collaboration~\cite{Ablikim:2017pyl}. It was shown
that final state interaction~\cite{Haidenbauer:2016won} or valence quark Coulomb enhancement factor~\cite{Baldini:2007qg}
are deficient in accounting for it. These anomalous behaviours are still waiting for a reasonable explanation.

As a matter of fact, new structures beyond the pQCD power-law behavior were seen in the data of $e^+e^-\to p\bar{p}$ by
BaBar collaboration~\cite{Lees:2013ebn}. It was found that these peaks are possibly related of to the resonances~\cite{Lorenz:2015pba}.
The most distinct peak would be the known $\phi$(2170) with the width of around 80 MeV observed by several experimental
groups (see~\cite{Lorenz:2015pba,Ablikim:2017auj,PDG:2018Tan} for a summary of data). The higher structure can be fitted
well with a broader vector meson at about 2.43 GeV. Though alternative explanations, e.g. thresholds opening~\cite{Lorenz:2015pba}
or large imaginary optical potential in the inelastic rescattering~\cite{Bianconi:2015owa,Bianconi:2015vva}, are not completely
excluded, these results give us enlightening insight into the $e^+e^-\to \Lambda\bar{\Lambda}$. In this paper, we are attempting
to investigate the role of vector mesons besides the usual pQCD contribution in this reaction.

\section{Formalism and results} \label{sec:resul}

Assuming that one photon exchange dominates the production, the Born cross section of baryon anti-baryon pairs through
virtual photon in $e^+e^-$ annihilation can be written in terms of the effective form factor $G_{\textrm{eff}}(s)$,
\be \label{eq:tcs}
\sigma_{e^+e^-\to \gamma^*\to B\bar{B}} (s) = \frac{4\,\pi\,\alpha^2\,\beta\, C}{3\,s} \lf(1 + \frac{1}{2\,\tau}\rg) |G_{\textrm{eff}}(s)|^2\,,
\ee
with $\tau = {s/4m^2}$ and $\beta = \sqrt{1 - 1/\tau}$. Here $\alpha\approx 1/137$ is the electromagnetic fine-structure
constant, $m$ the baryon mass, and $s$ the square of the center of mass (c.m.) energy. The Coulomb factor $C$ is accounting
for the electromagnetic interaction of the point-like baryon anti-baryon pairs in the final states~\cite{Baldini:2007qg}.
It is relevant to the close-to-threshold enhancement in $p\bar{p}$ final states but reads as 1 for the neutral $\Lambda\bar{\Lambda}$
interaction here.

The effective form factor can be generally expressed in terms of electric and magnetic form factors $G_E$ and $G_M$:
\be \label{eq:Geff}
|G_{\textrm{eff}}(s)| = \sqrt{\frac{2\,\tau\,|G_M(s)|^2 + |G_E(s)|^2}{1 + 2\,\tau}}\,,
\ee
which can be extracted from experimental measurements of $\sigma_{e^+e^-\to \gamma^*\to B\bar{B}} (s)$ by Eq.~(\ref{eq:tcs}). Based on the pQCD predictions~\cite{Lepage:1979za,Belitsky:2002kj},
the $|G_M|$ and $|G_E|$ in the time-like region can be parameterized as follows
\be \label{eq:GMpara}
  |G_M(s)| = |G_E(s)| = {\frac{A_B}{{\tau^2\ln^2(s/\Lambda^2_{QCD})}}} \,,\\
\ee
which are the same with the nucleon EFFs. Here $\Lambda_{QCD} = 0.3~\text{GeV}$ is the QCD scale parameter~\cite{Bianconi:2015owa,Bianconi:2015vva,Lorenz:2015pba}.
The only free parameter $A_B$ can be obtained by fitting
the experimental data. The factor $\ln(s/\Lambda^2_{QCD})$ represents the logarithmic corrections from QCD~\cite{Belitsky:2002kj},
which enables a good fit to the nucleon EFFs.

The form factor in Eq.~(\ref{eq:Geff}) is in fact only the non-resonant continuum associating with the direct $\gamma^*\to B\bar{B}$
transition. It does not consider the contribution of intermediate resonant states, e.g. $e^+e^-\to R_n \to B\bar{B}$.
The Fano-type form factor including the interference
between several resonances and one continuum background could be written as~\cite{Fano:1961zz}
\be \label{eq:FanoFFn}
|G_{\textrm{eff}}^\prime (s)|^2 = |G_{\textrm{eff}} (s)|^2 \frac{|\sum_n q_n \tan\Delta_n - 1|^2}{1 + (\sum_n \tan\Delta_n)^2} \,,
\ee
with $\tan\Delta_n = M_n \Gamma^{tot}_n /(s - M_n^2)$ for $n$-th resonance with mass $M_n$ and total width $\Gamma^{tot}_n$.
Obviously, $\Delta_n$ is related to the phase shift of $B\bar{B}$ scattering.
Eq.~(\ref{eq:FanoFFn}) corresponds to Eq.~(65) in Ref.~\cite{Fano:1961zz} with a bit alteration here.
When only one (n = 1) resonance is present, it reduces to be an usual form of one resonance and one continuum~\cite{Cao:2014qna,Wang:2014zha}
\be \label{eq:FanoFF1}
|G_{\textrm{eff}}^\prime (s)|^2 = |G_{\textrm{eff}} (s)|^2 \frac{|q + \varepsilon|^2}{1 + \varepsilon^2} \,,
\ee
with $\varepsilon = -\cot\Delta = (-s + M^2)/(M \Gamma^{tot})$. Eq.~(\ref{eq:FanoFF1}) is in fact minor modification of Eq.~(21) in Ref.~\cite{Fano:1961zz}.
In the present context, it is natural that the line shapes
of one state would not be the same in different channels, because the continuum would be different in various channels.
It has been demonstrated that this formula describes excellently the asymmetric line shape of the $\psi(3770)$ state, especially
the dip behind the resonant peak in the $e^+e^-\to D\bar{D}$ and $hadrons$~\cite{Cao:2014qna,Wang:2014zha}. One extraordinary
merit of the application of Fano-type formula to baryon EFFs is that we know very well the behaviour of the continuum $|G_{\textrm{eff}}(s)|$
from pQCD calculation. So the structures present in the data could be attributed to known or unknown resonances, which could be
conveniently added into the analysis by Eq.~(\ref{eq:FanoFFn}).

The Fano line shape parameter $q_n$ characterizes the electromagnetic transition probability of the resonant state. It
can be related to the relative transition amplitudes into the resonant state $T_{R_n}$ versus the continuum $T_{\gamma^*}$,
as clearly indicated by Eq.~(22) in Ref.~\cite{Fano:1961zz}:
\be \label{eq:modelq}
\frac{\pi}{2}\, q_n^2\, \Gamma^n_{B\bar{B}} = \frac{|<B\bar{B}|T_{\gamma^*}|e^+ e^->|^2}{|<B\bar{B}|T_{R_n}|e^+ e^->|^2}\,,
\ee
So the interference phase between two amplitudes is also contained in the parameter $q_n$.
Phenomenological model can be constructed to explicitly calculate these amplitudes, as in the case of $D\bar{D}$ reaction~\cite{Cao:2014qna,Wang:2014zha}.
Here the transition into continuum has been calculated by pQCD, but that into resonant states need the knowledge of the
inner structure of the corresponding states, which is beyond the scope of present paper. As can be seen from above equation,
$q_n$ is an energy dependent complex variable in the Fano scheme, but here we parameterize it as real for simplicity:
\be
q_n = \sqrt{\frac{\Gamma^{tot}}{\Gamma^n_{B\bar{B}}(s)}} q^0_n\,,
\ee
where $\Gamma^{tot}$ is introduced to make the $q_n$ dimensionless and $q^0_n$ is a constant determined by the data. The
$q_n$ could be regraded as a constant in the limited energy range, e.g.
for narrow state $\psi(3770)$ in $D\bar{D}$ reaction~\cite{Cao:2014qna,Wang:2014zha}.
We use the $s$-wave energy dependent width for below-threshold vector meson decaying to $\Lambda\bar{\Lambda}$,
\be
\Gamma_{\Lambda\bar{\Lambda}}(s) \propto p_{cm}(s) \propto \beta\sqrt{s} \,, 
\ee
with $p_{cm} = \sqrt{s/4 - m^2}$ being the final baryon momentum in the c.m. system. We use $\Gamma_{\Lambda\bar{\Lambda}}(s) = const.$
for the resonances above threshold.

\begin{figure}
  \begin{center}
  {\includegraphics*[width=9.0cm]{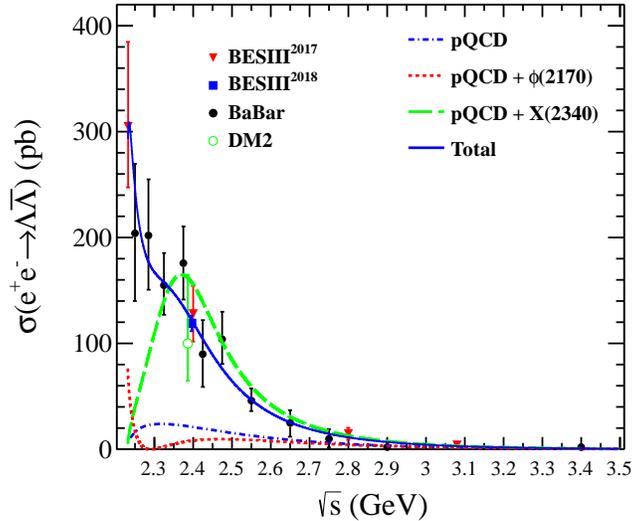}}
    \caption{(color online) Our fit to the world data of the cross section of $e^+e^-\to \Lambda\bar{\Lambda}$. The green open circle,
    black dot, red triangle and blue rectangle represent the data from the DM2~\cite{Bisello:1990rf}, BaBar~\cite{Aubert:2007uf}
    and BESIII~\cite{Ablikim:2017pyl,Cui:2018} collaborations, respectively. The blue dash-dotted curve describes the pQCD
    contribution. The red dotted and green dashed curves denote the contributions from pQCD involved with $\phi(2170)$ and
    $X(2340)$ resonances, respectively. The blue solid line is the total coherent contribution from pQCD, $\phi(2170)$, and
    $X(2340)$ resonances.
    \label{fig:tcsll}}
  \end{center}
\end{figure}

We find that two resonant mesons are needed to describe the available data of $e^+e^-\to \Lambda\bar{\Lambda}$. One of
them is the $\phi$(2170), which is required to explain the close-to-threshold enhancement. In view of the fact that this
state is below the $\Lambda\bar{\Lambda}$ threshold, its line shape is not fully present in the cross section of $e^+e^-\to
\Lambda\bar{\Lambda}$. It is impossible to unambiguously determine its properties here. We fix its mass and width to be 2.188 $\pm$ 0.010 GeV and
83.0 $\pm$ 12.0 MeV quoted from the Particle Data Group (PDG)~\cite{PDG:2018Tan}, respectively. Another one is a broad meson
with the mass of around 2.340 GeV (hereafter labeled as $X$(2340) for simplicity), which is demanded by the energy dependent behaviour of the cross section above threshold.
We let its mass and width be free parameters. The resulted curves are depicted in Figure~\ref{fig:tcsll}. It needs to be pointed
out that due to the larger uncertainty, the point marked by green open-circle from DM2 measurement~\cite{Bisello:1990rf}
is not included into the fit. A very good quality of agreement is achieved with 16 data points and five free parameters,
i.e. $\chi^2/\text{d.o.f} = 7.7/11 = 0.7$. The obtained numerical parameters are summarized in Table~\ref{tab:fit}. Here
the second uncertainties are associated with the mass and width of $\phi(2170)$ resonance, which are evaluated by
changing the mass and width by one standard deviation from the PDG values.
%

\begin{table}[t]
  \begin{center}
 \begin{tabular}{c|c|c|c|c}
\hline\hline
       $q^0_1$           &mass (GeV)               &width (MeV)       &$q^0_2$                &$A_\Lambda$\\
   $6.32\pm2.22\pm2.95$  &$2.338\pm0.046\pm0.030$  &$257\pm159\pm41$  &$-2.53\pm0.93\pm0.35$  &$1.19\pm0.46\pm0.09$\\
\hline\hline
   \end{tabular}
  \end{center}
  \caption{The resulted parameters in our fit. The second errors are obtained by varying the mass and width of $\phi(2170)$
  resonance by one standard deviation from the PDG values~\cite{PDG:2018Tan}.
  \label{tab:fit}}
\end{table}

As seen in Fig.~\ref{fig:tcsll}, the EFFs from the pQCD calculation contribute as a smooth background. The overall normalization
factor $A_\Lambda$ is smaller than the $A_p \sim 5.0$ for $e^+e^-\to p\bar{p}$~\cite{Ablikim:2014jrz}. The
ratio $A_\Lambda/A_p$ is roughly consistent with $(m_{u,d}/m_s)^2$ in the relativized constitute quark model~\cite{Godfrey:1985xj}.

We show the curves for total contribution of pQCD and one resonance by turning off another one by Eq.~(\ref{eq:FanoFFn}).
The $\phi$(2170) influences only the very close-to-threshold region, and it has a destructive interference with pQCD contribution.
The uncertainties of its mass and width only have small impact on the determination of pQCD contribution, as shown by the errors of $A_\Lambda$.
From Table~\ref{tab:fit}, we know that the $q_1^0$
for $\phi$(2170) has large uncertainty, reflecting the big error bars of the data at 2.2324 $\pm$ 0.00048 GeV, which is
1.0 MeV above the threshold. If the error of this energy point can be reduced or there are more precise measurements at different
near-threshold energies in the future, we can obtain more information about $\phi$(2170) and study its nature through model calculation
of its $q_1^0$ by Eq.~(\ref{eq:modelq}). The vector meson with the mass of around 2.340 GeV is broad, with a width of 257 $\pm$ 159 MeV. The large
uncertainty is originated from the big error bars of the data in this area. It provides the correct energy dependency of
the cross section over a wide energy range.

Whether these states in $e^+e^-\to \Lambda\bar{\Lambda}$ are related to the structures in $e^+e^-\to p\bar{p}$ is not clear
at present. The most distinct peak in $e^+e^-\to p\bar{p}$ is around 2.125 GeV, with a width of around 90 MeV~\cite{Lorenz:2015pba},
both compatible with the properties of $\phi$(2170) here. The position of the second peak is around 2.43 GeV in $e^+e^-\to p\bar{p}$,
which is higher than the second resonance here, though their widths are roughly close to each other~\cite{Lorenz:2015pba}.
No obvious structure is present in the data of $e^+e^-\to \Lambda\bar{\Lambda}$, which is different from the case of $e^+e^-\to p\bar{p}$.
This would be the reason that it is not easy to find the role of resonant states in $e^+e^-\to \Lambda\bar{\Lambda}$.
Our paper here point out the complicated interference of pQCD contribution, $\phi$(2170) and X(2340) in this reaction.
Because the $\Lambda$ hyperon also contains the strange quark, it is probable that the $\phi$(2170) and X(2340) states are more strongly coupled to $\Lambda\bar{\Lambda}$
than $p\bar{p}$. Therefore the resonant contribution shown in Fig. 1 are more clear than those in $e^+e^-\to p\bar{p}$ obtained in Fig. 5 and Fig. 6 of Ref.~\cite{Lorenz:2015pba}.
Also, $\phi$(2170) is above $p\bar{p}$ threshold but below $\Lambda\bar{\Lambda}$ threshold, so it distorts the cross sections differently in two reactions.
BESIII~\cite{Asner:2008nq} and Belle-II~\cite{Abe:2010gxa} collaborations have the chance to clarify these problems by measuring both $e^+e^-\to \Lambda\bar{\Lambda}$
and $e^+e^-\to p\bar{p}$ with higher precision and more statistics at more energy points in future.

\section{Discussion and Conclusion} \label{sec:conclusion}

In a brief summary, we investigate the role of vector mesons in the $e^+e^-\to \Lambda\bar{\Lambda}$ reaction. Our results support
the argument that the EFFs of $\Lambda$ hyperon obey the pQCD prediction. In other words, the deviation of the $e^+e^-\to
\Lambda\bar{\Lambda}$ to pQCD calculation is attributed to the contribution of two resonant mesons. We find that the $\phi$(2170)
is responsible for the close-to-threshold enhancement shown by the data, and another wide meson state with the mass around
2.34 GeV explains the energy dependency above threshold. If the structures in $e^+e^-\to p\bar{p}$ are confirmed by more
accurate experiment and really related to the vector mesons here, we could further determine some of their properties,
for example, separating their coupling to $e^+e^-$, $\Lambda\bar{\Lambda}$ and $p\bar{p}$ by a combined analysis of these reactions.

The nature of $\phi$(2170) state is widely studied in the literature~\cite{Chen:2018kuu,Dong:2017rmg,Zhao:2013ffn,Wang:2012wa,Chen:2011cj,MartinezTorres:2010ax,Coito:2009na,AlvarezRuso:2009xn,VaqueraAraujo:2009jq,Chen:2008ej,MartinezTorres:2008gy,Napsuciale:2007wp,Ding:2007pc,Wang:2006ri,Ding:2006ya}. The possible explanations include the strangeonium hybrid~\cite{Ding:2006ya,Ding:2007pc}, dynamical generated states~\cite{Napsuciale:2007wp,MartinezTorres:2008gy,MartinezTorres:2010ax,Coito:2009na,AlvarezRuso:2009xn,VaqueraAraujo:2009jq}, tetraquark~\cite{Wang:2006ri,Chen:2008ej,Chen:2018kuu} or $\Lambda\bar{\Lambda}$ molecular state~\cite{Zhao:2013ffn,Dong:2017rmg},
et al. Our finding that $\phi$(2170) state is important in the close-to-threshold $\Lambda\bar{\Lambda}$ production would
support its nature of $\Lambda \bar{\Lambda}$ molecule. Actually some evidence of $\Lambda\bar{\Lambda}$ threshold enhancement
has been also observed in the $J/\psi\to \gamma\Lambda\bar{\Lambda}$ decay~\cite{Ablikim:2012bw}, which is possibly associated
with the $\Lambda\bar{\Lambda}$ $^1$S$_0$ bound state~\cite{Zhao:2013ffn}.

There are several possible candidates for the higher lying state $X$(2340). We suggest that the $\omega$(2290)
with a mass of 2.290 $\pm$ 0.020 GeV and a width of 275 $\pm$ 35 MeV found in the partial wave analysis of
$p\bar{p}\to \Lambda\bar{\Lambda}$~\cite{Bugg:2004rj} is probably the same state in our fit here. Their values of mass and width are both highly overlapped with
the parameters in Table~\ref{tab:fit} within errors. Furthermore, they both couple strongly to the $\Lambda\bar{\Lambda}$. Among the further
light unflavored states listed by PDG~\cite{PDG:2018Tan}, another two $\omega$ states are also possible candidates with bigger spread of mass and width. They
are $\omega$(2330) with a mass of 2.330 $\pm$ 0.030 GeV and a width of 435 $\pm$ 75 MeV found in multi-pions production
in $\gamma p$ reaction\cite{Atkinson:1988xa}, and $\omega$(2205) with a mass of 2.205 $\pm$ 0.030 GeV and a width of 350 $\pm$ 90 MeV
seen in $p\bar{p}\to \omega\eta$ and $\omega\pi\pi$~\cite{PDG:2018Tan}. We also would like to point out that the recent data of multi-mesons production in $e^+e^-$
annihilation show a state with the mass of around 2.4 GeV and the width of around 100 MeV, with large uncertainties
due to the small statistical significance~\cite{Chen:2018kuu}. It would be the possible partner state of the $\phi$(2170)
as suggested in QCD sum rule~\cite{Chen:2018kuu}. Further study of $p\bar{p}$ annihilation by PANDA collaboration is definitely
welcome in order to drive a firm conclusion about the vector mesons in this energy range~\cite{Singh:2016dtf}. Moreover,
the goal of BESIII data taking~\cite{Asner:2008nq} is to accumulate 10 billion $J/\psi$ and 3 billion $\psi(3686)$ events,
thus these vector mesons can also be searched and studied in charmonium decays, such as $J/\psi, \psi(3686)\to \Lambda\bar{\Lambda}\pi^0/\eta$,
$\chi_{cJ}\to \Lambda\bar{\Lambda}\gamma/\phi/\omega$ and so on.

Our formalism and conclusion would give insight into the EMFFs of other baryons, e.g. $\Sigma$~\cite{Aubert:2007uf},
$\Lambda$(1520)~\cite{Ablikim:2018ckj} and $\Lambda_c^+$~\cite{Ablikim:2017lct,Pakhlova:2008vn} after they are measured with higher precision.
The simple fits in Ref.~\cite{Lorenz:2015pba} include the background terms with several effective pole terms below threshold and Breit-Wigner shapes that correspond to resonant states in order to describe the cross section. It is inapplicable to the near threshold enhancement in the data of $e^+e^-\to \Lambda\bar{\Lambda}$. Our formalism can give it a natural explanation, resulting into a satisfactory fit to $e^+e^-\to \Lambda\bar{\Lambda}$. Furthermore, it is based on more solid theoretical foundation, so it is possible to relate directly the fitting parameters to the properties of states if a phenomenological model considering the nature of states is constructed.

\begin{acknowledgments}

Useful discussions with Dr. Yu-tie Liang, and Professors Bing-Song Zou, Qiang Zhao, Yu-Bing Dong, Feng Yuan and Wen-Biao Yan are gratefully acknowledged. This work was supported by the National Natural Science Foundation of China (Grant Nos. 11405222 and 11505111).

\end{acknowledgments}

\end{document}